\documentstyle[12pt,epsfig]{article}
\def\lbldef#1#2{\expandafter\gdef\csname #1\endcsname {#2}}

\def\href#1#2{#2}  

\begin{document}
\baselineskip=15.5pt
\pagestyle{plain}
\setcounter{page}{1}

\begin{titlepage}

\begin{flushright}
CERN-TH/2000-058\\
hep-th/0002147
\end{flushright}
\vspace{10 mm}

\begin{center}
{\Large Bulk Fields in Dilatonic and Self-Tuning Flat Domain Walls}

\vspace{5mm}

\end{center}

\vspace{5 mm}

\begin{center}
{\large Donam Youm\footnote{E-mail: Donam.Youm@cern.ch}}

\vspace{3mm}

Theory Division, CERN, CH-1211, Geneva 23, Switzerland

\end{center}

\vspace{1cm}

\begin{center}
{\large Abstract}
\end{center}

\noindent

We study the Kaluza-Klein zero modes of massless bulk fields with various 
spins in the background of dilatonic and self-tuning flat domain walls.  We 
find that the zero modes of all the massless bulk fields in such domain wall 
backgrounds are normalizable, unlike those in the background of the 
non-dilatonic domain wall with infinite extra space of Randall and Sundrum.  
In particular,  gravity in the bulk of dilatonic domain walls is effectively 
compactified to the Einstein gravity with vanishing cosmological constant and 
nonzero gravitational constant in one lower dimensions for any values of 
dilaton coupling parameter, provided the warp factor is chosen to decrease 
on both sides of the domain wall, in which case the tension of the domain 
wall is positive.  However, unexpectedly, for the self-tuning flat domain 
walls, the cosmological constant of the zero mode effective gravity action  
in one lower dimensions does not vanish, indicating the need for additional 
ingredient or modification necessary in cancellation of the unexpected 
cosmological constant in the graviton zero mode effective action.

\vspace{1cm}
\begin{flushleft}
CERN-TH/2000-058\\
February, 2000
\end{flushleft}
\end{titlepage}
\newpage

\section{Introduction}

Recently, theories with extra spatial dimensions have been actively studied 
as solutions to the hierarchy problem in particle physics.  In such scenario, 
all the fields of Standard model are assumed to live within the worldvolume 
of a brane, whereas gravity can freely propagate in the extra space as well 
as within the worldvolume of the brane.   Although the earlier proposal 
\cite{ad1,ad2} attempts to solve the hierarchy problem by assuming large 
enough extra space, the hierarchy problem is recast into the problem of large 
ratio between the (fundamental) TeV Planck scale and the compactification 
scale of the extra spatial dimensions.  Later proposal \cite{ran1,ran2,ran3} 
by Randall and Sundrum (RS) relies on the exponentially decreasing warp 
factor in the metric of the non-factorizable spacetime for solving the 
hierarchy problem.  The RS model does not require large enough extra 
dimensions for solving the hierarchy problem; it is the decreasing warp 
factor that accounts for much smaller electroweak scale compared to the 
Planck scale.   Another novelty of the RS model is that the extra dimensions 
can even be infinite in size because gravity in the bulk of the RS domain 
wall is effectively localized \cite{ran2} around the domain wall.  

In our previous works \cite{youm1,youm2}, we showed that the RS type scenario 
can be extended to the dilatonic domain walls.  In fact, one is bound to 
consider dilatonic domain walls if one wishes to embed the RS type 
scenarios within string theories, since most of the five-dimensional 
domain walls in string theories are dilatonic.  We showed that the warp 
factor decreases \cite{youm1} (and becomes zero at finite distance away from 
the wall) and the gravity in the bulk of the dilatonic domain walls is 
effectively compactified to the Einstein gravity with vanishing cosmological 
constant and nonzero gravitational constant in one lower dimensions 
\cite{youm2}, if the cosmological constant is negative ($\Lambda=-2Q^2/\Delta
>0$, therefore $\Delta<0$, in our convention for the action).  In this paper, 
we show that even for the positive cosmological constant ($\Lambda<0$ in our 
convention and therefore $\Delta>0$) the bulk gravity is effectively 
compactified to the Einstein gravity with vanishing cosmological constant 
and nonzero gravitational constant in one lower dimensions, provided the warp 
factor is chosen to decrease on both sides of the wall and the extra 
spatial dimension is cut off through the introduction of another 
domain wall.  Note, for any values of the dilaton coupling parameter 
(therefore, for any values of $\Delta$), the warp factor can be chosen to be 
increasing or decreasing within the finite allowed extra spatial coordinate 
interval around the wall.  (In our previous works, we considered the case 
when the warp factor decreases [increases] for $\Delta<0$ [$\Delta>0$].)  
Therefore, the RS type scenario can be realized for dilatonic domain walls 
with any values of the dilaton coupling parameter, so for any dilatonic domain 
walls in string theories.  

The RS scenario relies on the fine-tuning of the domain wall tension, 
whose value is determined by the five-dimensional cosmological constant.  
It is pointed out \cite{beh} that the fine-tuned value of the domain 
wall tension is required by supersymmetry.  However, so far it has been 
observed \cite{beh,tow,gib,kall,pop,kal2,beh2} that supersymmetry rather 
requires increasing warp factor at least on one side of the wall, instead of 
decreasing warp factor on both sides.  Furthermore, it is not guaranteed that 
the fine-tuned value of the domain wall tension does not receive quantum 
corrections after the SUSY breaking.  Recently, new type of domain wall 
solutions which do not suffer from such problem were constructed 
\cite{adk,kss}.  These solutions, called ``self-tuning flat domain wall'', 
are obtained within the model with bulk dilaton but without bulk cosmological 
constant term.  Novelty of such domain wall solutions is that a static domain 
wall solution with the Poincar\'e invariance in one lower dimensions exists 
for any values of the domain wall tension.  So, even if the quantum effect 
corrects the domain wall tension, the Poincar\'e invariance is not disturbed.  
With a choice of the warp factor that has singularity at finite distance 
away from the wall on both sides, one therefore would expect that such 
domain wall effectively compactifies the five-dimensional gravity to 
four-dimensional gravity with vanishing cosmological constant 
regardless of the quantum corrections on the domain wall tension.  

It is the purpose of this paper to study the Kaluza-Klein (KK) zero modes 
of the massless fields with various spins in the bulk of the dilatonic and 
the self-tuning flat domain walls.   (The previous works on bulk fields 
in the non-dilatonic domain wall can be found for example in Refs. 
\cite{gw1,gw2,gw3,dhr,pom,bg,gn,chn}.)  Although the brane world scenarios 
assume that only gravity lives in the bulk and the remaining fields 
are confined within the brane worldvolume, it would be interesting to study 
various fields in the bulk.  One of the reasons is that if we want to 
embed the brane world scenarios within string theories, we have to consider 
bulk fields compactified from ten or eleven dimensions, unless we want to 
truncate them {\it ad hoc} just for the purpose of letting only gravity 
(and the dilaton for the dilatonic domain wall case) live in the bulk.  
We find that the KK zero modes of massless fields of various spins 
in the bulk of the dilatonic and the self-tuning flat domain walls are 
normalizable, whereas the KK zero modes of only the massless spin-0 and 
spin-2 fields in the bulk of non-dilatonic RS domain wall with infinite 
extra spatial dimensions \cite{ran2,ran3} are normalizable.  In general, we 
find that the KK zero modes for the integer spin bulk fields are independent 
of the extra spatial coordinate, whereas those of the half-integer spin fields 
depend on it.  An unexpected result is that, contrary to the claims in 
Refs. \cite{adk,kss}, the KK zero mode effective action (in one lower 
dimensions) for the gravity in the bulk of the self-tuning flat domain wall 
has non-vanishing cosmological constant term.  This seems to be an indication 
of the need for additional yet unknown ingredient in the picture of the 
self-tuning flat domain wall necessary in cancellation of the unexpected 
cosmological constant term in the effective action.

The paper is organized as follows.  In section 2, we discuss the dilatonic 
and the self-tuning flat domain walls.  We reparametrize the dilatonic 
domain walls, which we previously studied, in terms of the bulk cosmological 
constant.  We rederive the self-tuning flat domain walls with different 
parametrization (from the one in Ref. \cite{kss}) which we find more 
convenient.  In section 3, we obtain the KK zero modes for the bulk fields 
with various spins and their effective actions in one lower dimensions.

\section{Domain Wall Solutions}

In this section, we discuss the dilatonic domain walls that we studied 
previously and the self-tuning flat domain wall solutions that 
are constructed in Refs. \cite{adk,kss}.  Although detailed derivation is 
given in Ref. \cite{kss}, we rederive the self-tuning flat domain wall 
solutions because we wish to use different parametrization of the solutions, 
which we find more convenient.  Also, even if only five-dimensional domain 
walls are phenomenologically of interest, we derive the domain wall solutions 
in arbitrary dimensions just for the purpose of generality and because such 
solutions might be useful for other studies.

Generally, the total action is the sum of the $D$-dimensional action 
$S_{\rm bulk}$ in the bulk of the domain wall and the $(D-1)$-dimensional 
action $S_{\rm DW}$ on the domain wall worldvolume:
\begin{equation}
S=S_{\rm bulk}+S_{\rm DW}.
\label{totact}
\end{equation}
For the domain wall solutions under consideration in the paper, the 
$D$-dimensional action contains the bulk action for the domain wall solutions:
\begin{equation}
S_{\rm bulk}\supset {1\over{2\kappa^2_D}}\int d^Dx\sqrt{-G}\left[{\cal R}_G
-{4\over{D-2}}\partial_M\phi\partial^M\phi+e^{-2a\phi}\Lambda\right],
\label{blkdwsol}
\end{equation}
and the $(D-1)$-dimensional action contains the following worldvolume action 
for the domain wall solutions:
\begin{equation}
S_{\rm DW}\supset -\int d^{D-1}x\sqrt{-\gamma}f(\phi),
\label{wvdwsol}
\end{equation}
where $\gamma$ is the determinant of the induced metric $\gamma_{\mu\nu}=
\partial_{\mu}X^M\partial_{\nu}X^NG_{MN}$ on the domain wall worldvolume, 
$M,N=0,1,...,D-1$ and $\mu,\nu=0,1,...,D-2$.  Note, for the purpose of 
following the notation of our previous works \cite{youm1,youm2}, we are using 
the different convention for the sign of the cosmological constant $\Lambda$ 
and the sign and the value of the dilaton coupling parameter $a$ from that in 
Ref. \cite{kss}.  So, in our case, the positive [the negative] $\Lambda$ 
(with $a=0$) corresponds to the AdS space [dS space].  We are interested in 
finding solutions with the Poincar\' e invariance in $(D-1)$ dimensions.  The 
general Ans\" atze for the fields with such symmetry are
\begin{equation}
G_{MN}dx^Mdx^N=e^{2A(y)}\left[-dt^2+dx^2_1+\cdots+dx^2_{D-2}\right]+dy^2,
\label{genmetanz}
\end{equation}
and $\phi=\phi(y)$.  

The $\Lambda\neq 0\neq a$ case, considered in Ref. \cite{kss}, is nothing 
but the extreme dilatonic domain wall solutions studied in our previous works 
\cite{youm1,youm2}.  In this case, $f(\phi)$ in the worldvolume action has to 
take a specific form $f(\phi)=\sigma_{DW}e^{-a\phi}$ with the energy density 
$\sigma_{DW}$ of the domain walls taking the fine-tuned value determined 
by the bulk cosmological constant $\Lambda$ and the dilaton coupling 
parameter $a$.  It is observed in Ref. \cite{youm1} that the extra spatial 
coordinate $y$ terminates at finite nonzero value due to the singularity.  
It is explicitly shown in Ref. \cite{youm2} that the gravity in the 
bulk of the dilatonic domain wall is effectively compactified to the 
Einstein gravity in one lower dimensions with nonzero gravitational 
constant and vanishing cosmological constant.  The solution, reparametrized 
in terms of the bulk cosmological constant $\Lambda$ and the dilaton 
coupling parameter $a$, has the following form:
\begin{equation}
e^{2A}=(Ky+1)^{8\over{(D-2)^2a^2}},\ \ \ \ \ \ \ 
\phi={1\over a}\ln(Ky+1)+C,
\label{dildwsol}
\end{equation}
where $C$ is an integration constant and
\begin{equation}
K=\pm{{(D-2)a^2}\over 2}e^{-aC}\sqrt{{{D-2}\over{4(D-1)-a^2(D-2)^2}}\Lambda}.
\label{qval}
\end{equation}
The requirement of the term inside the square root in Eq. (\ref{qval}) to be 
positive fixes the sign of the cosmological constant $\Lambda$ to be positive 
[negative] for $a^2<4(D-1)/(D-2)^2$ [$a^2>4(D-1)/(D-2)^2$].  This is in 
accordance with the expression $\Lambda=-2Q^2/\Delta$, where $\Delta\equiv
(D-2)a^2/2-2(D-1)/(D-2)$, that we used in our previous works 
\cite{youm1,youm2}.  The sign $\pm$ in Eq. (\ref{qval}) has to be chosen such 
that the spacetime metric (\ref{genmetanz}) with the warp factor in Eq. 
(\ref{dildwsol}) has a singularity at finite $y$, if we want the gravity in 
the bulk of the domain wall to be effectively compactified.   So, we choose 
the plus [minus] sign for the region $y<0$ [$y>0$].  With this choice of the 
signs, the energy density of the wall, determined by solving the boundary 
condition at $y=0$, takes the following positive value:
\begin{equation}
\sigma_{\rm DW}={4\over{\kappa^2_D}}\sqrt{{{D-2}\over{4(D-1)-a^2(D-2)^2}}
\Lambda}.
\label{dwed}
\end{equation}
With a choice of the same signs on both sides of the wall, $\sigma_{DW}=0$.  
With a choice such that there is no singularity at $y\neq 0$, 
$\sigma_{DW}$ is negative with the same absolute value as Eq. (\ref{dwed}).  
These properties of the extreme dilatonic domain wall are essentially 
what we discussed in our previous works \cite{youm1,youm2}, except the choice 
of the sign $\pm$ in Eq. (\ref{qval}).  In our previous works, we considered 
the case when the metric has the singularity at finite $y$ for $\Delta<0$,  
but no singularity at finite $y$ for $\Delta>0$.  However, we see from 
the above that there is another choice of signs which gives the opposite 
singularity properties.  So, the correct statement has to be that {\it 
for any values of the dilaton coupling parameter $a$, one can choose 
the warp factor to be decreasing on both sides of the wall by choosing the 
sign $\pm$ in Eq. (\ref{qval}) such that $K<0$ [$K>0$] for $y>0$ [$y<0$], 
in which case $\sigma_{\rm DW}>0$ and the there are singularities on 
both sides of the wall}.  However, we note that for the $\Delta>0$ case (or 
the $a^2>4(D-1)/(D-2)^2$ case), the cosmological constant $\Lambda$ has the 
opposite sign, i.e., the bulk spacetime is dS-like.  

Now, we consider the $\Lambda=0$ case.  In deriving the solutions, we choose 
the static gauge for the domain wall worldvolume action, so $\gamma_{\mu\nu}
=\delta^M_{\mu}\delta^N_{\nu}G_{MN}$.  With the $(D-1)$-dimensional Poincar\'e 
invariant Ans\" atze for the fields, the equation of motion for the dilaton 
$\phi$ takes the following form:
\begin{equation}
{4\over{D-2}}{1\over{\kappa^2_D}}\left[(D-1)A^{\prime}\phi^{\prime}+
\phi^{\prime\prime}\right]=f^{\prime}(\phi)\delta(y),
\label{phieqn}
\end{equation}
and the Einstein's equations can be brought to the following forms:
\begin{equation}
{1\over 2}(D-2)(D-1)(A^{\prime})^2-{2\over{D-2}}(\phi^{\prime})^2=0,
\label{eineq1}
\end{equation}
\begin{equation}
{1\over{\kappa^2_D}}\left[(D-2)A^{\prime\prime}+{4\over{D-2}}
(\phi^{\prime})^2\right]+f(\phi)\delta(y)=0,
\label{eineq2}
\end{equation}
where the primes on $\phi$ and $A$ mean differentiations with respect to $y$ 
and the prime on $f(\phi)$ means differentiation with respect to $\phi$.  
Eq. (\ref{eineq1}) implies the following relation between $\phi$ and $A$:
\begin{equation}
A^{\prime}=\eta{2\over{D-2}}{1\over\sqrt{D-1}}\phi^{\prime}\ \ \ \ \ \ \ 
(\eta\equiv\pm 1).
\label{aphirel}
\end{equation}
So, Eqs. (\ref{phieqn}) and (\ref{eineq2}) can be brought to the following 
forms:
\begin{equation}
{4\over{D-2}}{1\over{\kappa^2_D}}\left[\eta{2\over{D-2}}\sqrt{D-1}
(\phi^{\prime})^2+\phi^{\prime\prime}\right]=f^{\prime}(\phi)\delta(y),
\label{pheq1}
\end{equation}
\begin{equation}
{1\over{\kappa^2_D}}\left[\eta{2\over\sqrt{D-1}}\phi^{\prime\prime}
+{4\over{D-2}}(\phi^{\prime})^2\right]=-f(\phi)\delta(y).
\label{pheq2}
\end{equation}

In the bulk ($y\neq 0$), Eqs. (\ref{pheq1}) and (\ref{pheq2}) take the same 
form and one can solve one of them to obtain the general solution for $\phi$.  
In Ref. \cite{kss}, the solution is parametrized in the following way:
\begin{equation}
\phi=\eta{{D-2}\over 2}{1\over\sqrt{D-1}}\ln|{2\over{D-2}}\sqrt{D-1}y+c|+d,
\label{oldpara}
\end{equation}
but we find it more convenient to parametrize the solution in the following 
way:
\begin{equation}
\phi=\eta{{D-2}\over 2}{1\over\sqrt{D-1}}\ln(Ky+1)+C.
\label{newpara}
\end{equation}  
Then, from Eq. (\ref{aphirel}) we obtain the following standard form of 
the metric warp factor:
\begin{equation}
e^{2A}=(Ky+1)^{2\over{D-1}}.
\label{warpfact}
\end{equation}
Note, Ref. \cite{kss} considers the possibility of having an arbitrary 
constant factor in front of this warp factor, which results from the 
integration constant in integrating Eq. (\ref{aphirel}) with respect to 
$y$ as well as from the integration constant $C$ in Eq. (\ref{newpara}).  
However, such constant factor can be absorbed by rescaling the 
$(D-1)$-dimensional coordinates $x^{\mu}$ in the metric.  The only effect 
of such constant factor on the KK zero modes of massless bulk fields is 
values of their normalization constants.  

Note, $\eta$'s in the above equations and the dilaton solution can take any 
(independent) signs on each side of the domain wall.  So, we have two 
sets of equations (\ref{pheq1}) and (\ref{pheq2}), one for each side, with 
$\eta$ replaced by $\eta_+$ [$\eta_-$] for the set of equations in the 
region $y>0$ [$y<0$].  And we have the following expressions for the 
warp factor:
\begin{equation}
e^{2A}=\left\{\matrix{(K_+y+1)^{2\over{D-1}},\ \ \ \ \ y>0\cr
(K_-y+1)^{2\over{D-1}},\ \ \ \ \ y<0}\right.,
\label{warppm}
\end{equation}
and the dilaton
\begin{equation}
\phi=\left\{\matrix{\eta_+{{D-2}\over 2}{1\over\sqrt{D-1}}\ln(K_+y+1)+C,
\ \ \ \ \ y>0\cr
\eta_-{{D-2}\over 2}{1\over\sqrt{D-1}}\ln(K_-y+1)+C,
\ \ \ \ \ y<0}\right.,
\label{dilpm}
\end{equation}
where we have imposed the continuity of $\phi$ at $y=0$, i.e., $\phi(0^+)=
\phi(0^-)$.  In the following, we impose the boundary conditions 
at $y=0$ on the general solutions (\ref{warppm}) and (\ref{dilpm}) to 
determine the integration constants $K_{\pm}$ in terms of the parameters 
in the actions (\ref{blkdwsol}) and (\ref{wvdwsol}) and the other integration 
constant.  

First, we consider the case when $\eta$'s in the above equations and the 
solutions on the two sides of the domain wall have the opposite signs.  
Namely, we choose $\eta_+=-\eta_-=\eta$, where $\eta=\pm 1$.  By imposing 
the boundary conditions on $\phi$ at $y=0$, we obtain the following 
expressions for $K_{\pm}$:
\begin{eqnarray}
K_+&=&{1\over 2}\kappa^2_D\left[\eta{\sqrt{D-1}\over 2}f^{\prime}(C)-
{{D-1}\over{D-2}}f(C)\right],
\cr
K_-&=&{1\over 2}\kappa^2_D\left[\eta{\sqrt{D-1}\over 2}f^{\prime}(C)+
{{D-1}\over{D-2}}f(C)\right].
\label{expofqpm}
\end{eqnarray}
With a choice $f(\phi)=\sigma_{\rm DW}e^{b\phi}$, $K_{\pm}$ 
take the following forms:
\begin{eqnarray}
K_+&=&{1\over 2}\kappa^2_D\sigma_{\rm DW}e^{bC}\left[\eta{\sqrt{D-1}\over 2}b-
{{D-1}\over{D-2}}\right],
\cr
K_-&=&{1\over 2}\kappa^2_D\sigma_{\rm DW}e^{bC}\left[\eta{\sqrt{D-1}\over 2}b+
{{D-1}\over{D-2}}\right].
\label{partqpm}
\end{eqnarray}
From these expressions for $K_{\pm}$, we see that a non-trivial solution 
does not exist when $b=\pm 2\sqrt{D-1}/(D-2)$.  As we will see in the 
following, non-trivial solutions for this case exist when $\eta$'s on the two 
sides of the wall have the same signs.  

Second, we consider the case when $\eta$'s on the two sides of the domain wall 
have the same signs.  Namely, we choose $\eta_+=\eta_-=\eta$, where $\eta=\pm 
1$.  By imposing the boundary conditions on $\phi$ at $y=0$, we obtain the 
following relation between $K_+$ and $K_-$:
\begin{equation}
K_+-K_-=\eta{\sqrt{D-1}\over 2}\kappa^2_Df^{\prime}(C)=-{{D-1}\over{D-2}}
\kappa^2_Df(C).
\label{qpmrel}
\end{equation}
With a choice $f(\phi)=\sigma_{\rm DW}e^{b\phi}$, the relation becomes:
\begin{equation}
K_+-K_-=\eta{\sqrt{D-1}\over 2}\kappa^2_Db\sigma_{\rm DW}e^{bC}
=-{{D-1}\over{D-2}}\kappa^2_D\sigma_{\rm DW}e^{bC},
\label{expqpmrel}
\end{equation}
from which we see that $b$ is fixed to take the following values:
\begin{equation}
b=-2\eta{\sqrt{D-1}\over{D-2}}.
\label{valofb}
\end{equation}
Note, we have freedom of choosing any values of $K_{\pm}$ as long as the 
constraint (\ref{qpmrel}) or (\ref{expqpmrel}) is satisfied.

We now comment on novelty \cite{adk,kss} of the domain wall solution with 
$\Lambda=0$.  First of all, as we can see from the explicit expressions 
for $K_{\pm}$, the $(D-1)$-dimensional Poincar\' e invariant solution exists 
for any values of the energy density $\sigma_{DW}$ of the domain wall (and 
the parameter $b$).  (In fact, the independent free parameters
\footnote{Another free parameter, namely the integration constant which 
results from integrating Eq. (\ref{aphirel}) with respect to $y$, can be 
removed by rescaling the $(D-1)$-dimensional coordinates $x^{\mu}$, as we 
mentioned previously.}, 
which can take any values, of the $\Lambda=0$ solutions are the parameters 
of $f(\phi)$, i.e., $\sigma_{DW}$ and $b$ for the $f(\phi)=\sigma_{DW}
e^{b\phi}$ case, and $C$.)  So, $\sigma_{DW}$ (and $b$) needs not be 
fine-tuned and the quantum correction on $\sigma_{DW}$ does not disturb the 
$(D-1)$-dimensional Poincar\' e invariance.  Second, in general $K_+\neq 
-K_-$, namely the solution has no ${\bf Z}_2$ invariance under $y\to -y$.  
The ${\bf Z}_2$ invariance is not possible for the $\eta_+=-\eta_-$ case, 
but one can choose the values of $K_{\pm}$ such that the ${\bf Z}_2$ 
invariance is achieved for the $\eta_+=\eta_-$ case.  So, the ${\bf Z}_2$ 
invariant domain wall solution obtained in Ref. \cite{adk} is a special case 
of the self-tuning flat domain wall solution with $\eta_+=\eta_-$ constructed 
in Ref. \cite{kss}.  All of these special features of the self-tuning flat 
domain wall solutions are manifestly recognizable, if we use our 
parametrization (\ref{newpara}) of the solutions, instead of the one 
(\ref{oldpara}) used in Ref. \cite{kss}.  Also, the solutions take simple 
and attractive forms with our parametrization. 

\section{The Kaluza-Klein Zero Modes of Bulk Fields}

In this section, we study the KK zero modes of massless fields in the 
bulk of various domain walls discussed in the previous section.  Although 
we are interested in the KK zero modes of massless bulk fields, we shall 
first obtain general equations satisfied by any KK modes of massive (for the 
spin-0 and spin-1 cases) bulk fields, and then we shall restrict ourselves 
to the special case of the KK zero modes of massless bulk fields.  In this 
section, we consider the five-dimensional domain walls, only, since only 
these are phenomenologically of interest.  So, in the following we rewrite 
the explicit domain wall solutions specifically for the $D=5$ case:
\begin{itemize}
\item dilatonic domain wall:\\
The warp factor and the dilaton for the domain wall with singularities 
at finite $y$ (therefore, decreasing warp factor) on both sides of the wall 
are given by
\begin{equation}
e^{2A}=(1-K|y|)^{8\over{9a^2}},\ \ \ \ \ \ \ \ \ \ 
\phi={1\over a}\ln(1-K|y|)+C,
\label{5dddw}
\end{equation}
where the parameter $K$ and the tension $\sigma_{\rm DW}$ of the wall 
take the following fixed values determined by the bulk cosmological constant 
$\Lambda$:
\begin{equation}
K={3\over 2}a^2e^{-aC}\sqrt{{3\over{16-9a^2}}\Lambda},\ \ \ \ \ 
\sigma_{\rm DW}={4\over{\kappa^2_5}}\sqrt{{3\over{16-9a^2}}
\Lambda}.
\label{5dpara}
\end{equation}

\item self-tuning flat domain wall:\\
The warp factor and the dilaton are given by
\begin{eqnarray}
e^{2A}&=&\left\{\matrix{(K_+y+1)^{1\over 2},\ \ \ \ \ y>0\cr
(K_-y+1)^{1\over 2},\ \ \ \ \ y<0}\right.,
\cr
\phi&=&\left\{\matrix{\eta_+{3\over 4}\ln(K_+y+1)+C,
\ \ \ \ \ y>0\cr
\eta_-{3\over 4}\ln(K_-y+1)+C,
\ \ \ \ \ y<0}\right..
\label{5dstfdw}
\end{eqnarray}
First, for the $\eta_+=-\eta_-=\eta$ case, the parameters $K_{\pm}$ are 
given in general by
\begin{eqnarray}
K_+&=&{1\over 2}\kappa^2_5\left[\eta f^{\prime}(C)-{4\over 3}f(C)\right],
\cr
K_-&=&{1\over 2}\kappa^2_5\left[\eta f^{\prime}(C)+{4\over 3}f(C)\right].
\label{5dexpofqpm}
\end{eqnarray}
With a choice $f(\phi)=\sigma_{\rm DW}e^{b\phi}$, 
\begin{eqnarray}
K_+&=&{1\over 2}\kappa^2_5\sigma_{\rm DW}e^{bC}\left[\eta b-{4\over 3}\right],
\cr
K_-&=&{1\over 2}\kappa^2_5\sigma_{\rm DW}e^{bC}\left[\eta b+{4\over 3}\right].
\label{5dpartqpm}
\end{eqnarray}
Second, for the $\eta_+=\eta_-=\eta$ case, the parameters $K_{\pm}$ are in 
general constrained to satisfy
\begin{equation}
K_+-K_-=\eta\kappa^2_5f^{\prime}(C)=-{4\over 3}\kappa^2_5f(C).
\label{5dqpmrel}
\end{equation}
With a choice $f(\phi)=\sigma_{\rm DW}e^{b\phi}$,
\begin{equation}
K_+-K_-=\eta\kappa^2_5b\sigma_{\rm DW}e^{bC}=-{4\over 3}\kappa^2_5
\sigma_{\rm DW}e^{bC}.
\label{5dexpqpmrel}
\end{equation}
\end{itemize}

\subsection{Scalar field}

The action for the bulk scalar field $\Phi(x^{\mu},y)$ is
\begin{equation}
S_{\rm bulk}\supset {1\over 2}\int d^4xdy\sqrt{-G}\left[G^{MN}\partial_M
\Phi\partial_N\Phi-m^2\Phi^2\right],
\label{scalact}
\end{equation}
where $G_{MN}$ is the metric (\ref{genmetanz}) for the domain wall solution.  
From this action, we obtain the following the equation of motion for the 
scalar:
\begin{equation}
\partial_M\left[\sqrt{-G}G^{MN}\partial_N\Phi\right]+\sqrt{-G}m^2\Phi=0,
\label{eomscal}
\end{equation}
which takes the following form after the metric (\ref{genmetanz}) is 
substituted:
\begin{equation}
e^{2A}\eta^{\mu\nu}\partial_{\mu}\partial_{\nu}\Phi+\partial_y[e^{4A}
\partial_y\Phi]+m^2e^{4A}\Phi=0.
\label{eomscal2}
\end{equation}

To consider the KK mode of $\Phi$ with mass $m_n$, only, we decompose 
$\Phi$ as 
\begin{equation}
\Phi(x^{\mu},y)=\varphi_n(x^{\mu})f_n(y),
\label{scaldec}
\end{equation}
and require $\varphi_n(x^{\mu})$ to satisfy the following Klein-Gordon 
equation for a scalar with mass $m_n$ in flat four-dimensional spacetime:
\begin{equation}
\left[\eta^{\mu\nu}\partial_{\mu}\partial_{\nu}+m^2_n\right]\varphi_n=0.
\label{mslsssc}
\end{equation}
Then, the equation of motion (\ref{eomscal2}) for $\Phi$ reduces to the 
following form of Sturm-Liouville equation:
\begin{equation}
\partial_y\left[e^{4A}\partial_yf_n\right]+m^2e^{4A}f_n=
m^2_ne^{2A}f_n.
\label{sceq1}
\end{equation}
The operator ${\cal L}=\partial_y(e^{4A}\partial_y)+m^2e^{4A}$ is 
self-adjoint, provided the boundary condition $\left.f^*_ne^{4A}f^{\prime}_m
\right|^{y=b}_{y=a}=0$ is satisfied, where $a\leq y\leq b$ is the interval 
in which the domain wall metric (\ref{genmetanz}) is well-defined.  In this 
case, the eigenvalues $m^2_n$ are real and the eigenfunctions $f_n$ with 
different eigenvalues are orthogonal to each other with respect to the 
weighting function $w(y)=e^{2A}$, i.e., $\int^a_bdyf^*_m(y)f_n(y)w(y)=0$ 
for $m^2_m\neq m^2_n$.

In term of a new $y$-dependent function $\tilde{f}_n=e^{2A}f_n$, Eq. 
(\ref{sceq1}) takes the following form of the Schr\"odinger equation:
\begin{equation}
-{{d^2\tilde{f}_n}\over{dy^2}}+V(y)\tilde{f}_n=m^2\tilde{f}_n.
\label{sceq2}
\end{equation}
with the potential
\begin{equation}
V(y)=2\left[A^{\prime\prime}+2(A^{\prime})^2+{1\over 2}m^2_ne^{-2A}\right].
\label{potsceq2}
\end{equation}
From now on, we shall be interested in only the zero mode ($m_0=0$) 
of the massless bulk scalar ($m=0$).  

First, we consider the dilatonic domain wall solution (\ref{5dddw}).  The 
potential (\ref{potsceq2}) in the Schr\"odinger equation (\ref{sceq2}) takes 
the following form:
\begin{equation}
V(y)={{8K^2}\over{81a^4}}{{8-9a^2}\over{(1-K|y|)^2}}-{{16K}\over{9a^2}}
\delta(y).
\label{potscddw}
\end{equation}
The ${\bf Z}_2$ invariant solution to the Schr\"odinger equation that 
satisfies the boundary condition $\tilde{f}^{\prime}_0(0^+)-
\tilde{f}^{\prime}_0(0^-)=-{{16K}\over{9a^2}}\tilde{f}_0(0)$ has the 
form $\tilde{f}_0(y)\sim (1-K|y|)^{8\over{9a^2}}$.  So, the KK zero mode 
is constant: $f_0(y)=e^{-2A}\tilde{f}_0(y)={\rm constant}$.
This constant zero mode is normalizable
\footnote{In the case of the non-dilatonic RS domain wall, the KK zero mode 
$f_0={\rm constant}$ is also normalizable: $\int^{\infty}_{-\infty}dy\,f^2_0
e^{2A}\sim \int^{\infty}_{-\infty}dy\,e^{-2k|y|}<\infty$.} 
with respect to the weighting 
function $w(y)=e^{2A}$.  The normalization constant is $N_0=\sqrt{({1\over 2}
+{4\over{9a^2}})K}$, i.e., $\int^{1/K}_{-1/K}dy\,f^*_0(y)f_0(y)w(y)=1$ with 
$f_0(y)=N_0$ and $w(y)=e^{2A}$.  Indeed, by substituting the KK zero-mode 
$\Phi=\varphi_0f_0$ into the action (\ref{scalact}) for the massless bulk 
scalar ($m=0$), we obtain the following action for the massless scalar in 
the flat four-dimensional spacetime:
\begin{eqnarray}
{1\over 2}\int d^4xdy\sqrt{-G}G^{MN}\partial_M\Phi\partial_N\Phi&=&
{1\over 2}\int^{1/K}_{-1/K}dy\,e^{2A}f^2_0\int dx^4\eta^{\mu\nu}
\partial_{\mu}\varphi_0\partial_{\nu}\varphi_0
\cr
&=&{1\over 2}\int dx^4\eta^{\mu\nu}\partial_{\mu}\varphi_0\partial_{\nu}
\varphi_0.
\label{scalact4d}
\end{eqnarray}
As we pointed out in the previous section, had we chosen to have a constant 
factor (due to the integration constants) in the warp factor $e^{2A}$, the 
normalization constant $N_0$ would have had dependence on the constant factor 
(as can be seen from the normalization relation $\int dyf^2_0e^{2A}=1$), but 
the effective action (\ref{scalact4d}) for the KK zero mode in one lower 
dimensions does not depend on the constant factor.  This generally holds 
for the case of the self-tuning flat domain walls and for other bulk fields 
to be discussed in the following subsections.  

Second, for the self-tuning flat domain wall solution (\ref{5dstfdw}), 
the potential takes the following form
\footnote{We define $A^{\prime}(y)$ at $y=0$ as $A^{\prime}(0)\equiv 
\lim_{\varepsilon\to 0^+}{{A(\varepsilon)-A(-\varepsilon)}\over
{\varepsilon-(-\varepsilon)}}$, and similarly for $A^{\prime\prime}(0)$.  
And we used the fact that $a+b\delta(y)=b\delta(y)$ at $y=0$ for any 
finite constant $a$.}:
\begin{equation}
V(y)=\left\{\matrix{-{1\over 4}{{K^2_+}\over{(K_+y+1)^2}}&,&\ \ \ \ y>0\cr
{{K_+-K_-}\over 2}\delta(y)&,&
\ \ \ \ y=0\cr
-{1\over 4}{{K^2_-}\over{(K_-y+1)^2}}&,&\ \ \ \ y<0}\right..
\label{potscstdw}
\end{equation}
The solution $f_0(y)$ to the Schr\"odinger equation satisfying the 
boundary condition $\tilde{f}^{\prime}_0(0^+)-\tilde{f}^{\prime}_0(0^-)=
{{K_+-K_-}\over 2}\tilde{f}_0(0)$ is $\tilde{f}_0(y)\sim (K_+y+1)^{1/2}$ 
for $y>0$ and $\sim (K_-y+1)^{1/2}$ for $y<0$.  So, as in the dilatonic 
domain wall case, the KK mode zero mode is constant: $f_0(y)=e^{-2A}
\tilde{f}_0(y)={\rm constant}$.  The normalized form of the zero mode 
is $f_0(y)=N_0=\sqrt{{3K_+K_-}\over{2(K_+-K_-)}}$.  Similarly as in 
the dilatonic domain wall case, one obtains the action for the massless 
dilaton $\varphi_0$ in the four-dimensional flat spacetime by substituting 
the zero mode field $\Phi=\varphi_0f_0$ into the bulk action (\ref{scalact}).  
In this case, the integration interval for the extra spatial coordinate is 
$-1/K_-\leq y\leq -1/K_+$.  

\subsection{Abelian gauge field}

The action for the bulk Abelian gauge field $A_M(x^{\mu},y)$ is
\begin{equation}
S_{\rm bulk}\supset \int dx^4dy\sqrt{-G}\left[-{1\over 4}G^{MN}G^{RS}F_{MR}
F_{NS}+{1\over 2}m^2G^{MN}A_MA_N\right],
\label{u1act}
\end{equation}
from which we obtain the following equation of motion for $A_M$:
\begin{equation}
{1\over\sqrt{-G}}\partial_M\left[\sqrt{-G}G^{MN}G^{RS}F_{NS}\right]+
m^2G^{RS}A_S=0.
\label{u1eq1}
\end{equation}
By taking the divergence (defined as $\nabla_MV^M\equiv{1\over\sqrt{-G}}
\partial_M(\sqrt{-G}V^M)$) of this equation, one has $m^2\nabla_MA^M=0$.  
For $m\neq 0$, one obtains the gauge condition $\nabla_MA^M=0$ on a massive 
$A_M$.  By using this gauge condition, one can eliminate one of the five 
components of $A_M$, which we choose as $A_y$.  Then, the gauge condition 
$\nabla_MA^M=0$ along with the gauge choice $A_y=0$ and the five-dimensional 
metric $G_{MN}$ of the form (\ref{genmetanz}) implies $\eta^{\mu\nu}
\partial_{\mu}A_{\nu}=0$.  In the case of massless ($m=0$) bulk Abelian gauge 
field $A_M$, one can also choose the gauge $A_y=0=\eta^{\mu\nu}\partial_{\mu}
A_{\nu}$ by using the gauge degrees of freedom.  In the $A_y=0=\eta^{\mu\nu}
\partial_{\mu}A_{\nu}$ gauge, the equation of motion (\ref{u1eq1}) for $A_M$ 
takes the following form:
\begin{equation}
\left[\eta^{\mu\nu}\partial_{\mu}\partial_{\nu}+\partial_ye^{2A}\partial_y+
m^2e^{2A}\right]A_{\rho}=0.
\label{u1eq2}
\end{equation}

To consider the KK mode of the bulk Abelian gauge field $A_{\rho}$ 
with mass $m_n$, only, we decompose $A_{\rho}$ as
\begin{equation}
A_{\rho}(x^{\mu},y)=a^{(n)}_{\rho}(x^{\mu})f_n(y),
\label{decomu1}
\end{equation} 
and require $a^{(n)}_{\rho}$ to satisfy the following Proca equation for an 
Abelian gauge field with mass $m_n$ in the Lorentz gauge in flat 
four-dimensional spacetime:
\begin{equation}
\left[\eta^{\mu\nu}\partial_{\mu}\partial_{\nu}+m^2_n\right]a^{(n)}_{\rho}=0.
\label{abelmasls}
\end{equation}
Then, the equation of motion (\ref{u1eq2}) reduces to the following 
Sturm-Liouville equation:
\begin{equation}
\partial_y\left[e^{2A}\partial_yf_n\right]+m^2e^{2A}f_n=m^2_nf_n.
\label{u1eq3}
\end{equation}
So, the KK modes with different masses are orthogonal to each other with 
respect to the weighting function $w(y)=1$, provided the boundary condition 
$\left.f^*_me^{2A}f^{\prime}_n\right|^{y=b}_{y=b}=0$ is satisfied.  

By using a new $y$-dependent function $\tilde{f}_n=e^Af_n$, one can 
bring Eq. (\ref{u1eq3}) to the following Schr\"odinger equation form:
\begin{equation}
-{{d^2\tilde{f}_n}\over{dy^2}}+V(y)\tilde{f}_n=m^2\tilde{f}_n.
\label{u1eq4}
\end{equation}
with the potential
\begin{equation}
V(y)=A^{\prime\prime}+(A^{\prime})^2+m^2_ne^{-2A}.
\label{potu1eq4}
\end{equation}
From now on, we consider only the zero mode ($m_0=0$) of the massless 
bulk Abelian gauge field ($m=0$).

First, for the dilatonic domain wall solution (\ref{5dddw}), the 
potential (\ref{potu1eq4}) in the Schr\"odinger equation (\ref{u1eq4}) 
takes the following form:
\begin{equation}
V(y)={{4K^2}\over{81a^4}}{{4-9a^2}\over{(1-K|y|)^2}}-{{8K}\over{9a^2}}
\delta(y).
\label{zmdpot}
\end{equation}
The solution to the Schr\"odinger equation that satisfies the 
boundary condition $\tilde{f}^{\prime}_0(0^+)-\tilde{f}^{\prime}_0(0^-)=
-{{8K}\over{9a^2}}\tilde{f}_0(0)$ is $\tilde{f}_0(y)\sim(1-K|y|)^{4\over
{9a^2}}$.  So, the KK zero mode is constant: $f_0(y)=e^{-A}\tilde{f}_0=
{\rm constant}$.  The normalized form of the zero mode is $f_0(y)=N_0=
\sqrt{K/2}$.  We obtain the following effective action for the massless 
Abelian gauge field $a^{(0)}_{\mu}$ (with the field strength $f^{(0)}_{\mu\nu}
=\partial_{\mu}a^{(0)}_{\nu}-\partial_{\nu}a^{(0)}_{\mu}$) in the 
four-dimensional flat spacetime by substituting the zero mode field 
$A_{\mu}=a^{(0)}_{\mu}f_0$ into the action (\ref{u1act}) for the bulk 
massless Abelian gauge field ($m=0$):
\begin{eqnarray}
-{1\over 4}\int dx^4dy\sqrt{-G}G^{MN}G^{RS}F_{MR}F_{NS}&=&
-{1\over 4}\int^{1/K}_{-1/K}dy\,f^2_0\int dx^4\eta^{\mu\nu}\eta^{\rho\sigma}
f^{(0)}_{\mu\rho}f^{(0)}_{\nu\sigma}
\cr
&=&-{1\over 4}\int dx^4\eta^{\mu\nu}\eta^{\rho\sigma}f^{(0)}_{\mu\rho}
f^{(0)}_{\nu\sigma}.
\label{u1act4d}
\end{eqnarray}

Second, for the self-tuning flat domain wall solution (\ref{5dstfdw}), 
the potential is given by
\begin{equation}
V(y)=\left\{\matrix{-{3\over{16}}{{K^2_+}\over{(K_+y+1)^2}}&,&\ \ \ \ \ y>0
\cr 
{{K_+-K_-}\over 4}\delta(y)&,&\ \ \ \ \ y=0
\cr 
-{3\over{16}}{{K^2_-}\over{(K_-y+1)^2}}&,&\ \ \ \ \ y<0}\right..
\label{potstdwabel}
\end{equation}
The solution $\tilde{f}_0(y)$ to the Schr\"odinger equation satisfying the 
boundary condition $\tilde{f}^{\prime}_0(0^+)-\tilde{f}^{\prime}_0(0^-)=
{{K_+-K_-}\over 4}\tilde{f}_0(0)$ is $\tilde{f}_0(y)\sim(1+K_+y)^{1/4}$ 
for $y>0$ and $\sim(1+K_-y)^{1/4}$ for $y<0$.  So, as in the dilatonic 
domain wall case, the KK zero mode is $y$-independent: $f_0(y)=e^{-A}
\tilde{f}_0={\rm constant}$.  The normalized form of the KK zero mode is 
$f_0(y)=N_0=\sqrt{K_+K_-/(K_+-K_-)}$.  Similarly as in the case of the 
dilatonic domain wall, we obtain the four-dimensional effective action 
for the massless Abelian gauge field $a^{(0)}_{\mu}$ in flat spacetime 
by plugging the zero mode field $A_{\mu}=a^{(0)}_{\mu}f_0$ into the 
bulk action. 

Note, in the case of the non-dilatonic domain wall of the original RS model 
\cite{ran1,ran2,ran3}, the KK zero mode $f_0={\rm constant}$ is not 
normalizable: $\int^{\infty}_{-\infty}dy\,f^2_0=\infty$ for the warp factor 
$e^{2A}=e^{-2k|y|}$ which is defined on $-\infty<y<\infty$.  So, the 
four-dimensional effective action for the massless Abelian gauge field cannot 
be obtained, unless one restricts the allowed values of $y$, for example, by 
regarding the extra dimension to be a segment $S^1/{\bf Z}_2$ as in the first 
RS model \cite{ran1}.  

\subsection{Spinor field}

The action for the bulk spin-1/2 fermion $\Psi(x^{\mu},y)$ is
\footnote{Note, for bulk fermions, we use the mostly negative metric signature 
convention, i.e., $\eta_{\mu\nu}={\rm diag}(1,-1,-1,-1,-1)$.}
\begin{equation}
S_{\rm bulk}\supset \int dx^4dy\sqrt{G}\,i\bar{\Psi}\Gamma^MD_M\Psi,
\label{fermact}
\end{equation}
where $D_M\equiv\partial_M+{1\over 4}\omega_{MAB}\gamma^{AB}$ is 
the gravitational covariant derivative on a spinor.  Here, $\omega_{MAB}$ 
is the usual spacetime spin-connection.  The convention for the 
spacetime vector indices are $M,N,P$ [$A,B,C$] for the five-dimensional 
curved [flat-tangent] spacetime indices, $\mu,\nu,\rho$ [$\alpha,\beta,
\gamma$] for the four-dimensional curved [flat-tangent] spacetime indices, 
and $y$ and 5 respectively for the curved and flat extra spatial indices.
The flat space gamma matrices $\gamma^A$ satisfying $\{\gamma^A,\gamma^B\}
=2\eta^{AB}$ are give by $\gamma^A=(\gamma^{\alpha},i\gamma^5)$.  The 
curved space gamma matrices $\Gamma^M\equiv E^M_A\gamma^A$ satisfy 
$\{\Gamma^M,\Gamma^N\}=2G^{MN}$, where $E^M_A$ is the inverse of the 
F\"unfbein $E^A_M$.  

From the above action, we obtain the following equation of motion for the 
bulk spinor $\Psi$:
\begin{equation}
i(\Gamma^M\partial_M+{1\over 4}\Gamma^M\omega_{MAB}\gamma^{AB})\Psi=0,
\label{fermeq1}
\end{equation}
which reduces to the following form after the metric (\ref{genmetanz}) is 
substituted:
\begin{equation}
i(e^{-A}\gamma^{\alpha}\partial_{\alpha}+i\gamma^5\partial_y+
2i\partial_yA\gamma^5)\Psi=0.
\label{fermeq2}
\end{equation}
To consider the KK mode with mass $m_n$, only, we decompose the bulk spinor 
$\Psi=\Psi^R+\Psi^L$ as $\Psi^{R,L}(x^{\mu},y)=\psi^{R,L}_n(x^{\mu})
f^{R,L}_n(y)$ and require $\psi_n=\psi^R_n+\psi^L_n$ to satisfy the following 
Dirac equation for a spinor with mass $m_n$ in flat four-dimensional spacetime:
\begin{equation}
\left[i\gamma^{\alpha}\partial_{\alpha}-m_n\right]\psi_n=0.
\label{fltdraceq}
\end{equation}
Here, $\Psi^{R,L}\equiv{1\over 2}(1\pm\gamma^5)\Psi$ and similarly for 
$\psi^{R,L}_n$.  Then, the equation of motion (\ref{fermeq2}) reduces to the 
following form:
\begin{equation}
\left(\partial_y+2\partial_yA\right)f^{R,L}_n=\pm m_ne^{-A}f^{L,R}_n.
\label{fermeq3}
\end{equation}
In the following, we study the KK zero mode ($m_0=0$).  

First, for the dilatonic domain wall solution (\ref{5dddw}), the KK 
zero mode is
\begin{equation}
f_0(y)\sim(1-K|y|)^{-{8\over{9a^2}}}.
\label{kkzmdfermddw}
\end{equation}
We check the normalizability of the KK zero mode by plugging the zero 
mode field $\Psi=\psi_0(x^{\mu})f_0(y)$ into the bulk action:
\begin{equation}
\int dx^4dy\sqrt{G}\,i\bar{\Psi}\Gamma^MD_M\Psi=
\int^{1/K}_{-1/K}dy\,e^{3A}f^2_0\int dx^4\,i\bar{\psi}_0\gamma^{\alpha}
\partial_{\alpha}\psi_0.
\label{fermact2}
\end{equation}
We see that the KK zero mode is normalizable, provided $a^2>4/9$.  The 
KK zero mode including the normalization factor is $f_0(y)=\sqrt{({1\over 2}-
{2\over{9a^2}})K}(1-K|y|)^{-{8\over{9a^2}}}$.  

Second, for the self-tuning flat domain wall solution (\ref{5dstfdw}), 
the KK zero mode is
\begin{equation}
f_0(y)\sim\left\{\matrix{(K_+y+1)^{-{1\over 2}},\ \ \ \ \ y>0\cr
(K_-y+1)^{-{1\over 2}},\ \ \ \ \ y<0}\right..
\label{kkzmdstdwfer}
\end{equation}
One can show that this KK zero mode is normalizable by plugging the zero 
mode field $\Psi=\psi_0(x^{\mu})f_0(y)$ into the bulk action, as we did for 
the dilatonic domain wall case.  The normalization factor for the KK zero mode 
in this case is $N_0=\sqrt{{3K_+K_-}\over{4(K_+-K_-)}}$.  

For the non-dilatonic RS domain wall, the zero mode $f_0(y)\sim e^{2k|y|}$, 
is not normalizable, i.e., $\int^{\infty}_{-\infty}dy\,e^{3A}f^2_0\sim 
\int^{\infty}_{-\infty}dy\,e^{k|y|}=\infty$, unless one restricts the 
allowed values of $y$ within a finite interval.  

\subsection{Gravitino}

The action for the bulk gravitino $\Psi_M$ is
\begin{equation}
S_{\rm bulk}\supset \int d^4xdy\sqrt{-G}\,{1\over 2}\bar{\Psi}_M
\Gamma^{MNP}D_N\Psi_P.
\label{gravnact}
\end{equation}
So, the equation of motion for the gravitino is
\begin{equation}
\Gamma^{MNP}D_N\Psi_P=0.
\label{gravteq}
\end{equation}
We choose the gauge $\Psi_y=0$.  

To consider the KK zero mode of $\Psi_{\mu}$, we decompose it as
\begin{equation}
\Psi_{\mu}(x^{\nu},y)=\psi^{(0)}_{\mu}(x^{\nu})f_0(y),
\label{gravtdec}
\end{equation}
and require $\psi^{(0)}_{\mu}$ to satisfy the following Rarita-Schwinger 
equation for the massless gravitino in flat four-dimensional spacetime:
\begin{equation}
\gamma^{\alpha\beta\delta}\partial_{\beta}\psi^{(0)}_{\delta}=0,
\label{mslsgtn4}
\end{equation}
along with the gauge conditions $\partial^{\alpha}\psi^{(0)}_{\alpha}=0=
\gamma^{\alpha}\psi^{(0)}_{\alpha}$.  Then, the equation of motion 
(\ref{gravteq}) takes the following form:
\begin{equation}
[\partial_y+\partial_yA]f_0=0.
\label{gravteq2}
\end{equation}

First, for the dilatonic domain wall solution (\ref{5dddw}), the KK zero 
mode is
\begin{equation}
f_0(y)\sim (1-K|y|)^{-{4\over{9a^2}}}.
\label{kkzmdddwrs}
\end{equation}
To check the normalizability of the KK zero mode, we substitute the zero 
mode field $\Psi_{\mu}=\psi^{(0)}_{\mu}f_0$ into the bulk action:
\begin{equation}
\int d^4xdy\sqrt{-G}\,{1\over 2}\bar{\Psi}_M\Gamma^{MNP}D_N\Psi_P=
\int^{1/K}_{-1/K}dy\,e^Af^2_0\int dx^4\,{1\over 2}\bar{\psi}^{(0)}_{\alpha}
\gamma^{\alpha\beta\delta}\partial_{\beta}\psi^{(0)}_{\delta}.
\label{gravnact2}
\end{equation}
We see that the KK zero mode (\ref{kkzmdddwrs}) is normalizable and its 
normalization factor is $N_0=\sqrt{K/2}$.  

Second, for the self-tuning flat domain wall solution (\ref{5dstfdw}), the 
KK zero mode is
\begin{equation}
f_0(y)\sim\left\{\matrix{(K_+y+1)^{-{1\over 4}},\ \ \ \ \ y>0\cr
(K_-y+1)^{-{1\over 4}},\ \ \ \ \ y<0}\right..
\label{kkzmdstwrs}
\end{equation}
Similarly as in the dilatonic domain wall case, one can show that this KK 
zero mode is normalizable.  The normalization factor of the KK zero 
mode $f_0(y)$, in this case, is $N_0=\sqrt{K_+K_-/(K_+-K_-)}$.  

For the non-dilatonic RS domain wall, the zero mode $f_0(y)\sim 
e^{k|y|}$ is not normalizable, i.e., $\int^{\infty}_{-\infty}dy\,e^Af^2_0
\sim \int^{\infty}_{-\infty}dy\,e^{k|y|}=\infty$, unless one restricts 
the allowed values of $y$ within a finite interval.  

\subsection{Graviton}

In our previous works, we observed \cite{youm1} that the RS type scenario 
can be extended to the extreme dilatonic domain walls and showed \cite{youm2} 
that indeed the dilatonic domain walls effectively compactify gravity to 
one lower dimensions by calculating the graviton KK zero mode effective 
action.  Later, this was further confirmed \cite{gjs} by explicitly 
constructing the normalizable graviton KK zero modes.  Although the explicit 
graviton KK zero modes are given in Ref. \cite{gjs}, we shall calculate them 
again since we are using different parametrization of solution, which we find 
to be more convenient.  

For the purpose of studying the KK zero mode of graviton, it is more 
convenient to consider the domain wall metric in conformally flat 
form.  (The reason is that in such coordinate frame the metric 
perturbation in the RS gauge around the domain wall metric takes 
the form of the Schr\"odinger equation (\ref{scheq}) in the below.)  This is 
achieved by transforming the extra spatial coordinate to new one, which 
we denote as $z$.  The conformal factor for the domain wall metric and the 
dilaton are
\footnote{Making use of the invariance of conformally flat form of the 
domain wall metric Ansatz under the translation in $z$, we choose the 
coordinate such that $y=0$ corresponds to $z=0$.  Then, the 
requirement ${\cal C}(z=0)=e^{2A(y=0)}=1$ fixes the conformal factor 
to take the following forms.}
\begin{equation}
{\cal C}(z)=(1+\bar{K}|z|)^{8\over{9a^2-4}},\ \ \ \ \ 
\phi={{9a}\over{9a^2-4}}\ln(1+\bar{K}|z|)+C,
\label{cfddw}
\end{equation}
for the dilatonic domain wall, and
\begin{eqnarray}
{\cal C}(z)&=&\left\{\matrix{(1+\bar{K}_+z)^{2\over 3},\ \ \ \ \ z>0\cr
(1+\bar{K}_-z)^{2\over 3},\ \ \ \ \ z<0}\right.,
\cr
\phi&=&\left\{\matrix{\eta_+\ln(1+\bar{K}_+z)+C,\ \ \ \ \ y>0\cr
\eta_-\ln(1+\bar{K}_-z)+C,\ \ \ \ \ y<0}\right.,
\label{cfsdw}
\end{eqnarray}
for the self-tuning flat domain wall, where $\bar{K}$ and 
$\bar{K}_{\pm}$ are defined as
\begin{equation}
\bar{K}\equiv{{4-9a^2}\over 6}e^{-aC}\sqrt{{3\over{16-9a^2}}\Lambda},
\ \ \ \ \ \ \ \ 
\bar{K}_{\pm}\equiv{3\over 4}K_{\pm}.
\label{kprimedef}
\end{equation}

To study the KK modes of graviton in the bulk background of the 
domain walls, we consider the following small fluctuation around the 
domain wall metric:
\begin{equation}
G_{MN}dx^Mdx^N={\cal C}(z)\left[(\eta_{\mu\nu}+h_{\mu\nu})dx^{\mu}dx^{\nu}
+dz^2\right],
\label{metpert}
\end{equation}
where metric perturbation $h_{\mu\nu}(x^{\rho},z)$ is assumed to satisfy 
the transverse traceless gauge condition $h^{\mu}_{\ \mu}=0=\partial^{\mu}
h_{\mu\nu}$.  The $(\mu,\nu)$-component of the Einstein equations is 
approximated, to the first order in $h_{\mu\nu}$, to
\begin{equation}
\left[\Box_x+\partial^2_z+{3\over 2}{{\partial_z{\cal C}}\over{\cal C}}
\partial_z\right]h_{\mu\nu}=0,
\label{1steineq}
\end{equation}
where $\Box_x\equiv\eta^{\mu\nu}\partial_{\mu}\partial_{\nu}$.  To 
consider the KK mode with mass $m_n$, only, we decompose $h_{\mu\nu}$ as 
$h_{\mu\nu}(x^{\rho},z)=\hat{h}^{(n)}_{\mu\nu}(x^{\rho})f_n(z)$ and require 
$\hat{h}^{(n)}_{\mu\nu}$ to satisfy $\Box_x\hat{h}^{(n)}_{\mu\nu}=m^2_n
\hat{h}^{(n)}_{\mu\nu}$.  Then, the linearized Einstein equation 
(\ref{1steineq}) reduces to the following form:
\begin{equation}
\left[\partial^2_z+{3\over 2}{{\partial_z{\cal C}}\over{\cal C}}
\partial_z+m^2_n
\right]f_n=0.
\label{1steineq2}
\end{equation}
Had we used $y$ as the extra spatial coordinate, the equation 
(\ref{1steineq2}) satisfied by the KK mode $f_n$ with mass $m_n$ would have 
taken the following form of the Sturm-Liouville equation:
\begin{equation}
\partial_y\left[e^{4A}\partial_yf_n\right]+m^2_ne^{2A}f_n=0,
\label{1steineq3}
\end{equation}
from which we know that the KK modes are orthogonalized with the 
respect to the weighting function $w(y)=e^{2A}$, provided the boundary 
condition $\left.f^*_me^{4A}f^{\prime}_n\right|^{y=b}_{y=a}=0$ is 
satisfied, or with respect to the weighting function $w(z)={\cal C}^{3/2}$ 
if $z$ is used as the extra spatial coordinate.   

In terms of a new $z$-dependent function defined as $\tilde{f}_n\equiv 
{\cal C}^{3/4}f_n$, Eq. (\ref{1steineq2}) takes the following form 
of the Schr\"odinger equation:
\begin{equation}
-{{d^2\tilde{f}_n}\over{dz^2}}+V(z)\tilde{f}_n=m^2_n\tilde{f}_n,
\label{scheq}
\end{equation}
with the potential
\begin{equation}
V(z)={3\over{16}}\left[4{{\cal C}^{\prime\prime}\over{\cal C}}-
\left({{\cal C}^{\prime}\over{\cal C}}\right)^2\right].
\label{schpot}
\end{equation}
In the following, we study the KK zero mode ($m_0=0$), for which 
$\hat{h}^{(0)}_{\mu\nu}$ satisfies the linearized vacuum Einstein equation 
$\Box_x\hat{h}^{(0)}_{\mu\nu}=0$ in the Lorentz gauge.

First, for the dilatonic domain wall solution (\ref{cfddw}), the 
potential (\ref{schpot}) in the Schr\"odinger equation (\ref{scheq}) 
takes the following form:
\begin{equation}
V(z)={{6\bar{K}^2}\over{(9a^2-4)^2}}{{10-9a^2}\over{(1+\bar{K}|z|)^2}}
-{{12\bar{K}}\over{4-9a^2}}\delta(z).
\label{zmdpotgrv}
\end{equation}
Note, the $\delta$-function potential is always attractive, implying that 
the KK zero mode solution can always be supported.  The solution 
$\tilde{f}_0$ to the Schr\"odinger equation satisfying the boundary 
condition $\tilde{f}^{\prime}_0(0^+)-\tilde{f}^{\prime}_0(0^-)=-{{12\bar{K}}
\over{4-9a^2}}\tilde{f}_0(0)$ is $\tilde{f}_0\sim(1+\bar{K}|z|)^{-{6\over
{4-9a^2}}}$.  So, the KK zero mode is $z$-independent: $f_0(z)={\cal 
C}^{-3/4}\tilde{f}_0={\rm constant}$.  The normalization factor for the KK 
zero mode is $N_0=\sqrt{({1\over 2}+{4\over{9a^2}})K}$.

Second, for the self-tuning flat domain wall solution (\ref{cfsdw}), 
the potential has the following form:
\begin{equation}
V(z)=\left\{\matrix{-{1\over 4}{{\bar{K}^2_+}\over{(\bar{K}_+z+1)^2}}
&,&\ \ \ \ \ z>0
\cr 
{1\over 2}(\bar{K}_+-\bar{K}_-)\delta(z)&,&\ \ \ \ \ z=0
\cr 
-{1\over 4}{{\bar{K}^2_-}\over{(\bar{K}_-z+1)^2}}&,&\ \ \ \ \ z<0}\right..
\label{potgvst}
\end{equation}
As in the dilatonic domain wall case, the $\delta$-function potential 
is always attractive.  The solution $\tilde{f}_0(z)$ to the Schr\"odinger 
equation satisfying the boundary condition $\tilde{f}^{\prime}_0(0^+)-
\tilde{f}^{\prime}_0(0^-)={1\over 2}(\bar{K}_+-\bar{K}_-)\tilde{f}_0(0)$ is 
$\tilde{f}_0(z)\sim (\bar{K}_+z+1)^{1/2}$ for $z>0$ and $\sim (\bar{K}_-z+
1)^{1/2}$ for $z<0$.  So, as in the dilatonic domain wall case, the KK zero 
mode is $z$-independent: $f_0(z)={\cal C}^{-3/4}\tilde{f}_0={\rm constant}$.  
The normalization constant for the KK zero mode $f_0$ is $N_0=
\sqrt{{3\over 2}{{K_+K_-}\over{K_+-K_-}}}$.

In the following, we obtain the KK zero mode effective actions for the 
graviton in the bulk backgrounds of the dilatonic and the self-tuning domain 
walls.  Since we have shown that the KK zero mode $h_{\mu\nu}=
\hat{h}^{(0)}_{\mu\nu}f_0$ of the graviton is independent of the extra spatial 
coordinate $z$, we consider the following form of the bulk metric:
\begin{equation}
G_{MN}dx^Mdx^N={\cal C}(z)\left[g_{\mu\nu}(x^{\rho})dx^{\mu}dx^{\nu}+dz^2
\right],
\label{bulkmet}
\end{equation}
where the conformal factor ${\cal C}$ is given by Eq. (\ref{cfddw}) or 
(\ref{cfsdw}).  The dilaton $\phi(z)$ is given by Eq. (\ref{cfddw}) or 
(\ref{cfsdw}).  A useful formula for obtaining the KK zero mode effective 
actions is
\begin{equation}
\sqrt{-G}{\cal R}_G=\sqrt{-g}{\cal C}^{3\over 2}\left[{\cal R}_g-4
{{{\cal C}^{\prime\prime}}\over{\cal C}}+\left({{{\cal C}^{\prime}}\over
{\cal C}}\right)^2\right].
\label{kkricci}
\end{equation}
Here, ${\cal R}_G$ and ${\cal R}_g$ are respectively the Ricci scalars for 
the metrics $G_{MN}$ and $g_{\mu\nu}$.  If we instead use the KK zero 
mode bulk metric of the following form:
\begin{equation}
G_{MN}dx^Mdx^N=e^{2A(y)}g_{\mu\nu}(x^{\rho})dx^{\mu}dx^{\nu}+dy^2,
\label{bulkwarpmet}
\end{equation}
then the corresponding formula would be
\begin{equation}
\sqrt{-G}{\cal R}_G=\sqrt{-g}\left[e^{2A}{\cal R}_g-4e^{4A}\left\{2
A^{\prime\prime}+5(A^{\prime})^2\right\}\right],
\label{kkricci2}
\end{equation}
but the resulting expression for the effective actions will be the same.

First, we consider the effective action in the bulk of the dilatonic 
domain wall.  In our previous work \cite{youm2}, we obtained the effective 
action for the case where $K$ defined in Eq. (\ref{qval}) is positive 
[negative] in the region $y<0$ [$y>0$] for the $\Delta<0$ case, 
only.  In this case, the tension $\sigma_{\rm DW}$ of the wall takes 
the positive value given by Eq. (\ref{dwed}) with $D=5$.  
In this paper, we shall assume that $K$ defined in Eq. (\ref{qval}) 
is positive [negative] in the region $y<0$ [$y>0$] even for the 
$\Delta>0$ case, as well as for the $\Delta<0$ case.  Substituting the  
Ans\"atze
\footnote{Such Ans\"atze are consistent with the five-dimensional equations 
of motion as long as the zero mode metric $g_{\mu\nu}(x^{\rho})$ satisfies 
the Ricci flat condition.  The Ricci flat condition is equivalent to 
the condition that the zero mode $g_{\mu\nu}(x^{\rho})$ satisfies the 
four-dimensional Einstein equations with vanishing cosmological constant.} 
for the fields given by Eqs. (\ref{cfddw}) and (\ref{bulkmet}) into 
the total action, we obtain the following:
\begin{eqnarray}
S&=&{1\over{2\kappa^2_5}}\int d^5x\sqrt{-G}\left[{\cal R}_G-{4\over 3}
\partial_M\phi\partial^M\phi+e^{-2a\phi}\Lambda\right]-\sigma_{DW}
\int d^4x\sqrt{-\gamma}e^{-a\phi}
\cr
&=&{1\over{2\kappa^2_5}}\int d^4xdz\sqrt{-g}\varpi^{{12}\over{9a^2-4}}\left[
{\cal R}_g-{{20(16-9a^2)\bar{K}^2}\over{(4-9a^2)^2}}\varpi^{-2}+
{{64\bar{K}}\over{4-9a^2}}\delta(z)\right.
\cr
& &\ \ \ \ \left.+e^{-2aC}\Lambda\varpi^{-2}\right]
-\int d^4x\sqrt{-g}e^{-aC}\sigma_{\rm DW},
\label{effactddwgr}
\end{eqnarray}
where from Eqs. (\ref{kprimedef}) and (\ref{5dpara}) we see that the 
cosmological constant and the tension of the wall can be expressed as
\begin{equation}
\Lambda={{12(16-9a^2)}\over{(4-9a^2)^2}}e^{2aC}\bar{K}^2,\ \ \ \ \ \ \ \ \ 
\sigma_{\rm DW}={1\over{\kappa^2_5}}{{24}\over{4-9a^2}}e^{aC}\bar{K},
\label{costenddw}
\end{equation}
and $\varpi\equiv 1+\bar{K}|z|$.  After the integration over $z$ (with the 
integration interval $-\infty<z<\infty$ for $a^2<4/9$ and $\bar{K}^{-1}<z<
-\bar{K}^{-1}$ for $4/9<a^2<16/9$), we see that all the extra terms cancel out 
and we are left with the term for the four-dimensional general relativity
(with vanishing cosmological constant) with the gravitational constant 
given by
\begin{equation}
\kappa^2_4={{8+9a^2}\over{2(4-9a^2)}}\bar{K}\kappa^2_5=e^{-aC}{{8+9a^2}\over
{12}}\sqrt{{{3}\over{16-9a^2}}\Lambda}\kappa^2_5.
\label{4dgravcon}
\end{equation}

A troublesome case is the $a^2>16/9$ case (with the integration interval 
$\bar{K}^{-1}<z<-\bar{K}^{-1}$), in which the integration on the $\varpi^{-2}$ 
terms in the action (\ref{effactddwgr}) diverges, whereas the effective 
four-dimensional gravitational constant $\kappa^2_4$ remains to have the 
nonzero value given by Eq. (\ref{4dgravcon}).  (This is the case which 
normally would have been discarded because of the positive cosmological 
constant (negative $\Lambda$ in our convention).)  On the other hand, 
the graviton KK zero mode $g_{\mu\nu}(x^{\rho})$ in the dilatonic domain wall 
background (\ref{cfddw}) satisfies the four-dimensional vacuum Einstein's 
equation with vanishing cosmological constant term.  So, it seems to be 
contradictory that we don't reproduce the action for the four-dimensional 
general relativity with vanishing cosmological constant by integrating the 
action (\ref{effactddwgr}) with respect to the extra spatial coordinate $z$.  
One can avoid infinity in the effective action by truncating the 
extra spatial dimension through the introduction of another domain 
wall in the region between $z=0$ and $|z|=-\bar{K}^{-1}$.  This seems 
to be reasonable because $|z|=-\bar{K}^{-1}$ (or $|y|=K^{-1}$) 
corresponds to the curvature singularity, which has to be avoided in 
the brane world scenario unless there is reasonable physical 
significance associated with the singularity.  Then, the total action 
has the following form:
\begin{eqnarray}
S&=&{1\over{2\kappa^2_5}}\int d^4x\int^{z_0}_{-z_0}dz\sqrt{-G}
\left[{\cal R}_G-{4\over 3}\partial_M\phi\partial^M\phi+e^{-2a\phi}
\Lambda\right]
\cr
& &-\sigma_{h}\int_{z=0} d^4x\sqrt{-\gamma^h}e^{-a\phi_h}
-\sigma_{v}\int_{z=z_0} d^4x\sqrt{-\gamma^v}e^{-a\phi_v},
\label{effactddwgr2}
\end{eqnarray}
where
\begin{eqnarray}
\sigma_{h}&=&-\sigma_{v}={4\over{\kappa^2_5}}\sqrt{{3\over{16-9a^2}}
\Lambda},
\cr
\gamma^h_{\mu\nu}&=&g_{\mu\nu},\ \ \ \ \ \ 
\gamma^v_{\mu\nu}=(1+\bar{K}z_0)^{8\over{9a^2-4}}g_{\mu\nu},
\cr
\phi_h&=&C,\ \ \ \ \ \ \ \ \ 
\phi_v={{9a}\over{9a^2-4}}\ln(1+\bar{K}z_0)+C.
\label{dwtens}
\end{eqnarray}
Going through the similar calculation as in Eq. (\ref{effactddwgr}), 
except that there will be an additional $\delta$-function term at $|z|=z_0$ 
in the integrand of the bulk action, one will find after the $z$-integration 
that all the extra terms are cancelled and one is left only with the curvature 
term.  The effective four-dimensional gravitational constant in this case is 
given by
\begin{eqnarray}
\kappa^2_4&=&{{8+9a^2}\over{2(4-9a^2)}}\bar{K}\left[1-\varpi^{{9a^2+8}\over
{9a^2-4}}_0\right]^{-1}\kappa^2_5
\cr
&=&e^{-aC}{{8+9a^2}\over{12}}\sqrt{{{3}\over{16-9a^2}}\Lambda}\left[1
-\varpi^{{9a^2+8}\over{9a^2-4}}_0\right]^{-1}\kappa^2_5,
\label{gravcon}
\end{eqnarray}
where $\varpi_0=1+\bar{K}z_0$.  Note, even if we introduced the 
additional domain wall to cure the problem of diverging effective 
action for the $a^2>16/9$ case, this result with the additional domain 
wall is valid for any values of $a$.  

So, we see that gravity in the bulk of the dilatonic domain wall is 
effectively compactified to one lower dimensions, for any values of the 
dilaton coupling parameter $a$, provided the extra space is truncated 
through the introduction of another domain wall in the case of 
$a^2>16/9$.  For this to happen, one has to choose the sign $\pm$ in the 
expression for $K$ in Eq. (\ref{qval}) such that the warp factor for the 
domain wall metric has singularity at a finite value of $y$ on both sides 
of the wall and therefore the warp factor decreases on both sides of the 
domain wall.  This choice of signs corresponds to the conformal factor of 
the form (\ref{cfddw}) (for the $D=5$ case) when the metric tensor is 
transformed to take a conformally flat form.  Also, in this case, the tension 
of the wall takes the positive value given by Eq. (\ref{dwed}).  For other 
choices of sign $\pm$ in the expression for $K$ in Eq. (\ref{qval}), 
the gravity in the bulk is not effectively compactified to the Einstein 
gravity with zero cosmological constant in one lower dimensions and 
the tension of the wall is either zero or negative.

Next, we consider the effective action in the bulk of the self-tuning flat 
domain wall.   Substituting the Ans\"atze for the fields given 
by Eqs. (\ref{cfsdw}) and (\ref{bulkmet}) into the total action, we obtain 
the following:
\begin{eqnarray}
S&=&{1\over{2\kappa^2_5}}\int d^5x\sqrt{-G}\left[{\cal R}_G-{4\over 3}
\partial_M\phi\partial^M\phi\right]-\int d^4x\sqrt{-\gamma}f(\phi)
\cr
&=&{1\over{2\kappa^2_5}}\int d^4xdz\sqrt{-g}\varpi_{\pm}\left[{\cal R}_g-
{8\over 3}(\bar{K}_+-\bar{K}_-)\delta(z)\right]-\int d^4x\sqrt{-g}f(C),
\label{effactstdwgr}
\end{eqnarray}
where in the second line we choose the plus [the minus] sign in $\varpi_{\pm}
\equiv 1+\bar{K}_{\pm}z$ for the $z>0$ [$z<0$] region and $f(C)=(\bar{K}_--
\bar{K}_+)/\kappa^2_5$.  Note, a term coming from ${\cal R}_G$ which 
would potentially have diverged after the $z$-integration is cancelled 
by another diverging term $-{4\over 3}\partial_M\phi\partial^M\phi$ and 
we are left only with the $\delta$-function term coming from ${\cal R}_G$.  
After the integration over $z$ (with the integration 
interval $-\bar{K}^{-1}_-<z<-\bar{K}^{-1}_+$), we obtain the following 
effective action:
\begin{equation}
S={1\over{2\kappa^2_5}}\int d^4x\sqrt{-g}{{\bar{K}_+-\bar{K}_-}\over
{2\bar{K}_+\bar{K}_-}}\left[{\cal R}_g-{4\over 3}\bar{K}_+\bar{K}_-\right].
\label{effactstdw}
\end{equation}
Unexpectedly, the four-dimensional effective action for the graviton zero 
mode has a cosmological constant term.  This is contradictory, because the 
graviton KK zero mode $g_{\mu\nu}(x^{\rho})$ in the bulk of the self-tuning 
flat domain wall satisfies the four-dimensional vacuum Einstein 
equations with vanishing cosmological constant and the effective 
four-dimensional gravitational constant has a nonzero value given 
by $\kappa^2_4={{2\bar{K}_+\bar{K}_-}\over{\bar{K}_+-\bar{K}_-}}\kappa^2_5$.  
It is also inconsistent with the fact that the self-tuning flat domain 
wall solution has the four-dimensional Poincar\'e invariance, which 
requires zero cosmological constant in four dimensions
\footnote{I would like to thank Prof. Kallosh for pointing out inconsistency 
between the existence of the domain wall solution with the four-dimensional 
Poincar\'e invariance and the non-vanishing cosmological constant in the 
effective action, after the first version of this paper appeared in the 
preprint archive.}.  
Ref. \cite{fll}, which appeared in the preprint archive after the present 
work, independently observes such unexpected nonzero cosmological 
constant term in the effective action and resolves the problem by 
introducing extra domain walls at the naked singularities
\footnote{The author feels that the extra introduced domain walls should 
rather be placed between the singularities and the self-tuning flat 
domain wall, because the singularities of the self-tuning flat domain 
wall are problematic regions where the induced metric vanishes, the dilaton 
blows up and the spacetime curvature diverges.} 
and {\it fine-tuning} their tensions to cancel out the undesirable 
cosmological constant term.  So, such resolution spoils nice self-tuning 
property of the self-tuning flat domain walls.  This seems to indicate 
that there is some other ingredient missing or some inconsistency in the 
scenario of self-tuning flat domain wall necessary in reproducing 
four-dimensional effective theory with vanishing cosmological constant 
without loosing nice self-tuning property.  (It seems to be important to 
obtain the expect form of the effective action, because anyway in the RS 
model we usually consider a part of the complete effective action to 
determine the four-dimensional effective gravitational constant.)  It is 
important to note that the existence of a solution with the Poincar\'e 
invariance in one lower dimensions and the fact that the graviton zero mode 
in the bulk of such solution satisfies the four-dimensional vacuum Einstein 
equation with zero cosmological constant do not necessarily mean that the 
effective action in one lower dimensions has the right expected form.  An 
example is the non-dilatonic domain wall with exponentially increasing warp 
factor.  The graviton zero mode in such background satisfies the 
four-dimensional vacuum Einstein equations with zero cosmological constant, 
but the four-dimensional gravitational constant is zero if we take the extra 
spatial dimension to be infinite in size.  Additional condition that the 
four-dimensional gravitational constant should be nonzero seems to be also 
insufficient.  As we have seen the case of the dilatonic domain walls with 
positive cosmological constant ($\Lambda<0$ in our convention), although the 
four-dimensional gravitational constant is nonzero, as well as the KK zero 
mode graviton in the domain wall bulk satisfies the four-dimensional vacuum 
Einstein equations with zero cosmological constant, the effective action 
diverges unless additional domain wall is introduced.  One possible solution 
to the problem might be to consider all the massive KK modes of bulk graviton 
to cancel out the unwanted cosmological constant, but on the other hand the 
graviton zero mode itself is a consistent solution to the five-dimensional 
equations of motion provided the zero mode $g_{\mu\nu}(x^{\rho})$ satisfies 
the Ricci flat condition.

\end{document}